# Comparison between Grating Imaging and Transient Grating Techniques on Measuring Carrier Diffusion in Semiconductor

*Ke Chen[1], Xianghai Meng[1], Feng He[1,2], Yongjian Zhou[1], Jihoon Jeong[1], Nathanial Sheehan[3], Seth R Bank[3], and Yaguo Wang[1,2]\**

\*Corresponding Author. Email: yaguo.wang@austin.utexas.edu.

1. Department of Mechanical Engineering, The University of Texas at Austin, Austin, TX, 78712, USA
2. Texas Materials Institute, The University of Texas at Austin, Austin, TX, 78712, USA
3. Department of Electrical and Computer Engineering, The University of Texas at Austin, Austin, TX, 78758, USA

Email：

Ke Chen: kechentp@gmail.com

Xianghai Meng: mengxiangh27@gmail.com

Feng He: hefeng89@gmail.com

Yongjian Zhou: Zhouyj33@gmail.com

Jihoon Jeong: jihoonjeong@utexas.edu

Nathanial Sheehan: nathanial.sheehan@gmail.com

Seth R Bank: sbank@utexas.edu

ORCid: 0000-0001-7646-8193

**ABSTRACT**

Optical grating technique, where optical gratings are generated via light inference, has been widely used to measure charge carrier and phonon transport in semiconductors. In this paper, compared are three types of transient optical grating techniques: transient grating diffraction, transient grating heterodyne, and grating imaging, by utilizing them to measure carrier diffusion coefficient in a GaAs/AlAs superlattice. Theoretical models are constructed for each technique to extract the carrier diffusion coefficient, and the results from all three techniques are consistent. Our main findings are: (1) the transient transmission change $\Delta T/T_0$ obtained from transient grating heterodyne and grating imaging techniques are identical, even these two techniques originate from different detection principles; and (2) By adopting detection of transmission change (heterodyne amplification) instead of pure diffraction, the grating imaging technique (transient grating heterodyne) has overwhelming advantage in signal intensity than the transient grating diffraction, with a signal intensity ratio of 315:1 (157:1).

**Keywords**: Transient grating heterodyne, grating imaging, diffraction, carrier diffusion

## I. INTRODUCTION

Carrier diffusion in semiconductors is crucial in electronic and opto-electronic devices, since it determines some key parameters of the devices, such as working frequency and response time. Studying carrier diffusion process can also reveal carrier scattering in semiconductors, assess carrier mobility with Einstein relation, and understand interactions between carriers and phonons, defects, and nanostructures. Currently, there are several optical techniques to measure the carrier diffusion coefficients nondestructively: transient grating [1,2], spatial scanning pump-probe [3,4], and grating imaging [5,6]. In the transient grating method, two pump beams overlap on the sample surface to generate a transient carrier density grating. A probe beam shines on the grating and the diffracted probe is taken as the signal, which reflects the decaying process of the carrier density grating. In the spatial scanning pump-probe technique, both the pump and probe beams are tightly focused onto the sample surface. The pump generates a Gaussian-shape carrier package and the probe is scanned spatially across the pump spot. By measuring the differential transmission or reflection ($\Delta T/T_0$ or $\Delta R/R_0$) of the probe as a function of time and position, the evolution of the carrier package, which contains the information of carrier diffusion, is recorded. In the grating imaging technique, pump and probe beams overlap on a physical transmission amplitude grating (a photomask with metal strips patterned onto a glass substrate), whose image is formed by an objective lens onto the surface of the sample. The intensities of pump and probe beam on the sample are modulated in the same pattern as the transmission amplitude grating. The pump generates transient carrier grating in the sample, while the probe only detects the evolution of carrier density in the bright-strip regions. By measuring either $\Delta T/T_0$ or $\Delta R/R_0$ of the probe as a function of time, the decay of the excited carrier density due to recombination and diffusion is monitored, from which the carrier diffusion coefficient can be extracted.

Among these techniques, the scanning pump-probe method provides more information since it directly detects both spatial and temporal evolution of the carrier diffusion process. However, this technique is time consuming because of the necessity to perform spatial scanning. When fast measurements are required, transient grating and grating imaging methods are preferred. One disadvantage of transient grating is the small diffraction efficiency that leads to a very weak signal.

In order to overcome this problem, heterodyne technique, where a reference beam is introduced to be collinear with the diffracted probe beam, has been implemented to amplify the diffraction signal [7,8]. The much stronger reference beam interferes with the diffracted probe beam and amplify the detected signal. However, optical alignment to achieve the spatial overlap between the reference beam and the diffracted probe beam used to be a challenge. Maznev et al [9] developed a novel optical setup where the requirement of spatial overlap between reference and probe beams is automatically satisfied, which popularizes transient grating heterodyne to measure various diffusion processes [10-12].

In this article, three optical grating techniques are revisited and compared: grating imaging, transient grating diffraction and transient grating heterodyne, by utilizing them to measure the in-plane carrier diffusion coefficient in a GaAs/AlAs superlattice. Theoretical models are also established for each technique. Our results show that the intensities of the raw signals of the grating image (transient grating heterodyne) technique is about 315 (157) times larger than that of the transient grating diffraction. It is also demonstrated that, in spite of the difference in the experimental setup, grating imaging and transient grating heterodyne techniques are actually interchangeable. They are two equivalent measurements from two different experimental perspectives.

## II. EXPERIMENTAL SETUPS OF DIFFERENT GRATING TECHNIQUES

In Figure 1, the schematic experimental setup of grating imaging method is shown, from which the setup of transient grating heterodyne and transient grating diffraction can be derived. After the transmission amplitude grating (photomask), pump and probe beams are diffracted into several beams. Three orders (0, and ±1) are collected by the objective lens and focused onto the sample surface. The positions of the transmission grating and the sample are carefully aligned to be conjugated, i.e. the real image of the grating is formed onto the sample surface. The intensities of the pump and probe beams are modulated into the same grating pattern, with alternating bright and dark fringes, as shown in Figure 1. The pump grating will generate carrier density grating in the sample, and the probe is only sensitive to the carrier density in the bright fringes. The carrier population in the bright fringes will decay due to both carrier recombination and carrier diffusion into the dark fringes. The differential transmission or reflection ($\Delta T/T_0$ or $\Delta R/R_0$) signals can reveal this decay process and provide the information about both carrier recombination and carrier diffusion. Without the transmission amplitude grating, the setup shown in Figure 1 returns to a traditional pump-probe spectrometer with unmodulated and relatively large laser spot on the sample surface (typically tens of µm), and the characteristic decay mainly reveals carrier cooling and recombination. With the carrier cooling and recombination times predetermined in the non-grating case, the carrier diffusion effect measured with grating imaging technique can be isolated to extract diffusion coefficients accurately.

With some simple modifications in the experimental setup (with blocker and selection of the detected beams), measurements for transient grating heterodyne or transient grating diffraction can be realized easily. Five cases are presented in Figure 2, all in side view. Case 1 is the grating imaging technique with 3 pump and 3 probes beams. Case 2 is the grating imaging technique with 2 pump and 2 probe beams. Case 3 and case 4 are the transient grating heterodyne with different directions (detection or reference) collected (to avoid confusion, after the objective lens, the up-pointing beam is referred as the reference beam and the down-pointing beam as detection beam), and case 5 is the transient grating diffraction geometry. In case 5, after the transmission grating, two orders (0 and -1) of the probe and the $0^{th}$ order of the pump are blocked, so that the transmitted +1 and -1 orders of the pump can still form transient carrier grating on the sample surface. Only one order (+1) of the probe beams is transmitted, and the diffraction of this beam is collected as signal. All these 5 setups are used to measure the carrier diffusion in a GaAs/AlAs superlattice.

The sample under test is a 30-periods GaAs (6nm)/AlAs (6nm) superlattice on a glass substrate. The superlattice was originally grown on GaAs substrate and then wet-etched and transferred onto a glass substrate. Details of sample growth and characterization can be found in Ref.[1]. The laser pulses come from a Ti: Sapphire oscillator with 80 MHz repetition rate and 100 fs pulse width (Spectra Physics, Tsunami). The laser spot size (for both pump and probe) on the sample surface is about 70 μm in diameter ($1/e^2$). The laser wavelength is 800 nm, resonant with the PL peak of our sample. Under resonant excitation, the differential transmission signal ($\Delta T/T_0$ or $\Delta R/R_0$) is predominated by the absorption coefficient change of the sample [13,14], which reflects the population of the excited carrier density governed by Pauli blocking effect. The period of the transmission amplitude grating (photomask) is 80 μm, with 40 μm opaque slit (metal strip) and 40 μm transparent slit. With 20X objective lens and according to optical interference principle, the period of the grating image formed on the sample surface is 2μm, with 1 μm bright and dark fringes, except for case 1, which has 4 μm period with 2 μm bright and dark fringes. The μm-order fringe width ensures that the transport of the excited carriers is due to carrier diffusion through multi scattering events, such as carrier-phonon scattering and carrier defect-scattering, since the electron mean free path in a semiconductor is typically on the order of tens of nanometers [15].

**III. EXPERIMENTAL RESULTS OF DIFFERENT GRATING TECHNIQUES**

Before measuring the sample with optical grating techniques, pump-probe experiments without amplitude grating are performed in order to: (1) make sure the system response to the laser excitation is in the linear region, e.g. $\Delta T/T_0$ or $\Delta R/R_0$ signals proportional with the excited carrier density [5,14]; (2) obtain the characteristic decay time from carrier cooling ($\tau_{cool}$) and carrier recombination ($\tau_r$), so that the pure carrier diffusion effect could be isolated in the grating imaging measurements. Figure 3a shows experimental $\Delta T/T_0$ signals measured at three different pump powers. Peak $\Delta T/T_0$ values (inset of Figure 3a) are proportional to the pump power. After normalization, as shown in Figure 3b, all three curves overlap, indicating not only the excitation but also the later part of the signal (carrier relaxation) is proportional to the pump power. This means that below 30 mW pump power, the response of the system to the excitation is in the linear region and the $\Delta T/T_0$ signals are proportional to the excited carrier density. The carrier relaxation process consists of two components: a fast one attributed to the carrier cooling effect and a slow one to carrier recombination. Thus the signal without grating can be described by the following expression: $\frac{\Delta T}{T_0}(t)_{non-grating} = cN(t) = cN_0(A_1 e^{-\frac{t}{\tau_{cool}}} + A_2 e^{-\frac{t}{\tau_r}})$, where $c$ is a proportional factor, $N$ and $N_0$ are the time-dependent excited carrier density and initial excited carrier density, $A_1$ and $A_2$ are the weight factors of carrier cooling and carrier recombination, and $\tau_{cool}$ and $\tau_r$ are the carrier cooling time and carrier recombination lifetime, respectively. In the grating imaging experiments, the pump power is kept below 30 mW to make sure all measurements are performed in the linear region.

Figure 4 shows the normalized $\Delta T/T_0$ signals with gratings of different cases presented in Figure 2, along with the non-grating case. Several features in Figure 4 are worth of notice. **(a)** Compared with the non-grating case, the signals with grating decay faster. This is due to the carrier diffusion effect: carrier diffusing away to the un-detected region (dark fringes in Fig.1) accelerates the decrease of the carrier density in the probed region (bright fringes in Fig.1). **(b)** Compared with case 1, the signal of case2 decays faster at the beginning but slower at longer time delay. According to optical interference principle, the spatial frequency of the carrier density grating generated with two pump beams (±1 orders) in case 2 is twice of that generated with three pump beams (0 and ±1 orders) in case 1. With the same initial carrier density in the bright fringes (equivalent to signal normalization), the effective carrier density gradient is larger in case 2 than in case 1. Therefore, the diffusion effect is stronger in case 2 than in case 1, which explains the faster decay for case 2 at the beginning. At the later time, carrier density grating will disappear and there will be only carrier recombination effect. As will be seen later in the derivation of theoretical models for both case 1 and case 2, when carrier density grating vanishes (no diffusion effect), the normalized signal will converge to a certain value, and this value is higher in case 2 than in case 1, which explain the slower decay at longer delay time in case 2. **(c)** The signal of case 2 overlaps with those of case 3 and case 4. Figure 2 shows that case 3 and case 4 are actually symmetric since the +1 order (reference beam) and -1 order (the detection beam) have the same intensity and phase. Case 2 can be viewed as a combination of case 3 and case 4, i.e., detecting both +1 and -1 order. Compared with case 3 (or 4), both $\Delta T$ and $T_0$ double in case 2, hence $\Delta T/T_0$ remains the same. However, by nature case 2 is the grating imaging setup with two pump and two probe beams, while case 3 and 4 are the transient grating heterodyne with (reference+detection diffraction) and (detection+reference diffraction) detected, respectively. The overlapping of the $\Delta T/T_0$ signals of cases 2 and case 3 (or 4) suggest that the grating imaging technique and the transient grating heterodyne technique are actually the same in terms of experimental results, even though with different principles of measurement. In the derivation of theoretical models, it is further demonstrated that the final analytical expressions of the detected signals for cases 2, 3 & 4 are equivalent, in spite of the fact that the derivations start from different perspectives. **(d)** The signal of case 5 decays the fastest with weak signal to noise ratio (SNR). In case 5, only the diffracted probe by the carrier grating is detected and the signal is only sensitive to the carrier diffusion, not to the carrier recombination. Once the carrier density grating vanishes and the carrier diffusion process stops, the diffraction signal will disappear. Therefore, the signal of case 5 shown in Figure 4 indicates that the diffusion process actually finishes at around 300 ps. The diffraction signal usually is extremely weak with amplitude comparable to the noise of the measurement system [1]. The comparison between case 5 and case 3 & 4 directly shows the substantial improvement in the SNR with the heterodyne technique (signal amplified by introducing a reference beam). Note that the signal of case 5 has even been averaged for 10 scans, but only 2 scans for case 3 & 4.

Figure 5 shows the absolute signal values for cases 2, 3 and 5, from the reading of an lock-in amplifier (LIA), under the same pump and probe power before the amplitude grating (pump/probe power ratio is 12/1). For case 2, the LIA reading just records the transmission change ΔT. For case 3, the LIA reading represents the diffraction and the reference heterodyned signal. For case 5, the LIA only reads the intensity of diffracted probe, $T_{diff}$. Figure 5 reveals the difference of the raw signal intensity among grating imaging, grating heterodyne and grating diffraction techniques. The intensity of transmission change in case 2 is about 315 times larger than the diffraction intensity in case 5. Considering there are 2 probe beams in case 2 while only one in case 5, the effective intensity of transmission change of a grating-patterned probe should be about 157 (315/2) times larger than the intensity of the diffraction of a plane-wave beam with identical power. Our result agrees with a previous report where transmission change was found 200 times larger than the diffraction intensity in a four-wave mixing measurement [16]. Thus, in terms of signal intensity, imaging grating technique is significantly advantaged by collecting transmission change, instead of diffraction. Figure 5 also shows that the signal intensity of the reference-heterodyned diffraction in case 3 is just half of the transmission change ΔT in case 2, which is reasonable since the detected beam in case 3 is just one of the two symmetrically detected beams in case 2, as discussed before. Thus, the signal intensity of the transient grating heterodyne geometry is also about 157 times larger than the pure diffraction intensity.

## IV. THEORETICAL MODELS TO EXTRACT DIFFUSION COEFFICIENTS FROM DIFFERENT GRATING TECHNIQUES

In order to extract the carrier diffusion coefficient from measurements with the various experimental geometries presented in Figure 2, what derived are the theoretical models for all the cases. It will be shown that the measurement principles of the grating imaging method and the transient grating heterodyne method can be understood from two perspectives.

**Case 1: Grating imaging with three pump beams and three probe beams**

In the grating imaging method, the transmission (or reflection) change of the probe beams is detected, and it is assumed that the local transmission (or reflection) change is proportional to the local excited carrier density. Such assumption is typically valid for small pump excitation power at resonant wavelength [13,14], as demonstrated in Figure 3. The local carrier density of the detected region will decrease due to carrier diffusion and recombination. The decrease will be reflected in the local transmission (or reflection) change. The measured entire transmission (or reflection) change is an integrated value over all the grating fringes. Therefore, the core idea of the grating imaging method is to collect the local transmission change over all the bright fringes that contain the information of carrier diffusion. Following the above measurement principle, the final analytical expression of the signal in case 1 is: (details of the derivation can be found in Ref. [1])

$$\Delta T(t) = \left(A_1 e^{-\frac{t}{\tau_r}} + A_2 e^{-\frac{t}{\tau_{cool}}}\right)\left[\left(\frac{1}{4}+\frac{2}{\pi^2}\right)^2 + \frac{2}{\pi^2}e^{-f^2 Dt} + \frac{2}{\pi^4}e^{-4f^2 Dt}\right] \quad (1)$$

where $f = 2\pi/P$ is the spatial frequency, P is the spatial grating period, $D$ is the carrier diffusion coefficient, $A_1$ and $A_2$ are the amplitudes for carrier cooling and carrier recombination, and $\tau_{cool}$ and $\tau_r$ are the carrier cooling time and carrier recombination lifetime, respectively. The first term in the square bracket of Equation (1) is a non-diffusion term, describing the weight of non-diffusion effect (carrier cooling and recombination), while the second and the third terms are diffusion-related, which will vanish at long time delay $t$, indicating that the diffusion effect will eventually vanish when the carrier density gradient decreases to zero. According to Equation (1), if the carrier lifetime $\tau_r$ is much larger than $1/f^2 D$ (which is our case, the fitted $\tau_r$=6648 ps and $1/f^2 D$=137 ps), when the diffusion effect vanishes, the remaining signal will approach to a certain value, which is $A_1 \left(\frac{1}{4}+\frac{2}{\pi^2}\right)^2$. And the ratio of the remaining signal over the initial signal is $A_1 \left(\frac{1}{4}+\frac{2}{\pi^2}\right)^2 / \left\{(A_1 + A_2)\left[\left(\frac{1}{4}+\frac{2}{\pi^2}\right)^2 + \frac{2}{\pi^2} + \frac{2}{\pi^4}\right]\right\} \approx 0.479 A_1/(A_1 + A_2)$. This ratio can be viewed as the percentage of the non-diffusion effect in the total signal. The ratio will be compared with that of case 2 to explain the difference observed in experimental data at longer time delay.

**Case 2: Grating imaging with two pump beams and two probe beams**

The derivation is similar to that of case 1. Right after pump excitation (time zero), two pump beams (+1 and -1 orders) generates a sinusoidal form of carrier density grating in the sample: $N(t = 0) = N_0 \left(\frac{1}{2}+\frac{1}{2}\cos(fx)\right)$, where $f = 2\pi/P$ is the spatial frequency, and P is the spatial period. The evolution of the carrier grating is governed by the carrier diffusion equation: $\frac{\partial N(t,x)}{\partial t} = D\nabla^2 N(t,x) - \frac{N(t,x)}{\tau_r}$, where $\tau_r$ is the carrier lifetime and $D$ is the diffusion coefficient. The solution of the equation is: $N(t,x) = N_0 e^{-\frac{t}{\tau_r}}\left(\frac{1}{2} + \frac{1}{2}\cos(fx)e^{-f^2Dt}\right)$. The local transmission intensity change $\Delta I(t,x,y)$ of probe is proportional to the local incident intensity $I_0(x,y)$ and the local carrier density: $\Delta I(t,x,y) = CI_0(x,y)N(t,x) = CI_{00}G(x,y)\left(\frac{1}{2} + \frac{1}{2}\cos(fx)\right)N(t,x)$, where $I_{00}$ is the pump incident intensity at the spot center, $G(x,y)$ is the spatial Gaussian profile of probe beam and $C$ is a proportional factor. The entire transmission change $\Delta T$ is the integral of all the local intensity change over all the grating periods: $\Delta T = \iint_{-\infty}^{+\infty} dxdy \Delta I(t,x,y) \approx Ae^{-\frac{t}{\tau_r}}(2 + e^{-f^2Dt})$, where $A = \frac{\pi r^2}{32 \ln 2} CI_{00}N_0$ is a factor. If further considering the carrier cooling, the final analytical expression of the signal in case 2 is:

$$\Delta T(t) = 2F(A_1 e^{-\frac{t}{\tau_r}} + A_2 e^{-\frac{t}{\tau_{cool}}})(1 + \frac{1}{2}e^{-f^2Dt}) \qquad (2)$$

Similar to case 1, the first term (the constant 1) in the second bracket on the right hand side is the non-diffusion term and the second term is the diffusion term. According to Equation (2), when the diffusion effect vanishes at long time delay, the ratio of the remaining signal over the initial signal is $A_1/\left[(A_1 + A_2)(1 + \frac{1}{2})\right] \approx 0.667 A_1/(A_1 + A_2)$, which is larger than the ratio in case 1 ($0.479 A_1/(A_1 + A_2)$ as estimated above). This explains why at later time delay when there is no diffusion effect, the signal of case 2 is higher than that of case 1, as shown in Figure 4.

**Case 3: Heterodyne of transmitted detection beam and diffracted reference beam**

In the transient grating heterodyne method, the generated carrier density grating modulates the refractive index in the sample and diffracts the incident detection beam. The electrical field of the output light (after passing through the sample), is determined by the product of the incident electrical field and a transmission (or reflection) transfer function modulated by the carrier density grating. With the specially designed heterodyne geometry, the diffracted detection beam and the transmitted reference beam will coincide in space automatically, interfere with each other, and be collected by the photodetector. Due to the additional reference beam, the detected signal also reflects both the carrier diffusion (from the diffracted detection beam) and recombination (from the transmitted reference beam) processes. The reference beam amplifies the diffraction signal through heterodyne interference. By monitoring the diffracted probe intensity, the grating heterodyne method gains information of carrier diffusion from the decay process of the carrier density grating. In principle, grating imaging method is transmission based, while the grating heterodyne is still diffraction based but with remarkable signal amplification.

Similar to case 2, two pump beams (+1 and -1 orders) generates a sinusoidal form of carrier density grating in the sample, $N(t = 0) = N_0 \left(\frac{1}{2} + \frac{1}{2}\cos(fx)\right)$. With the same governing equation for carrier density, the solution of carrier density grating is: $N(t,x) = N_0 e^{-\frac{t}{\tau_r}} \left(\frac{1}{2} + \frac{1}{2}\cos(fx) e^{-f^2 Dt}\right)$. The generated carrier grating will modulate the refractive index (mainly the imaginary part for detection at resonant wavelength, see supplemental material) [13,17], and the local modulation is proportional to the local excited carrier density for small excitation pump power. Thus the transmission transfer function is expressed as: $Trans(t,x) = \widetilde{t_0} + \Delta\tilde{t} = \widetilde{t_0} + c\,N(t,x)$, where $\widetilde{t_0}$ is the complex amplitude transmission coefficient for the unexcited state, $\Delta\tilde{t}$ is the transmission coefficient change induced by the excited carrier, and c is a proportional factor [10]. The optical fields of the incident reference beam and the incident detection beam are expressed as: $U_{ref-in} = E e^{ik_z z} e^{i\frac{f}{2}x}$ and $U_{det-in} = E e^{ik_z z} e^{-i\frac{f}{2}x}$, respectively. Thus, the output optical field of the detection beam and the reference beam after the sample is expressed as:

$$U_{ref-out} = U_{ref-in} Trans(t,x)$$

$$= Ee^{ik_z z}\left\{\left[\tilde{t}_0 + \frac{cN_0}{2}e^{-\frac{t}{\tau_r}}\right]e^{i\frac{f}{2}x} + \frac{cN_0}{4}e^{-\frac{t}{\tau_r}}e^{-f^2 Dt}e^{-i\frac{f}{2}x} + \frac{cN_0}{4}e^{-\frac{t}{\tau_r}}e^{-f^2 Dt}e^{i\frac{3f}{2}x}\right\}$$

and $U_{det-out} = U_{det-in} Trans(t,x)$

$$= Ee^{ik_z z}\left\{\left[\tilde{t}_0 + \frac{cN_0}{2}e^{-\frac{t}{\tau_r}}\right]e^{-i\frac{f}{2}x} + \frac{cN_0}{4}e^{-\frac{t}{\tau_r}}e^{-f^2 Dt}e^{i\frac{f}{2}x} + \frac{cN_0}{4}e^{-\frac{t}{\tau_r}}e^{-f^2 Dt}e^{-i\frac{3f}{2}x}\right\}$$

, respectively, where $N_0$ is the initial excited carrier density. The output of the reference beams consists of three directions, with

$k_x = \frac{f}{2}$ (transmitted, 0th order), $-\frac{f}{2}$ (diffracted, $-1$ order), $and$ $\frac{3f}{2}$ (diffracted, $+1$ order),

while the output of the detection beam also consists of three directions, with

$k_x = \frac{f}{2}$ (diffracted, $+1$ order), $-\frac{f}{2}$ (transmitted, 0 order), $and$ $-\frac{3f}{2}$ (diffracted, $-1$ order). It can be seen that the diffracted detection (reference) beam and the transmitted reference (detection) beam automatically coincide, both in the $k_x = \frac{f}{2}$ ($-\frac{f}{2}$) direction. Please note that the diffracted beams after the sample with $k_x = \frac{3f}{2}$ and $k_x = -\frac{3f}{2}$ are not shown in Figure 2 for simplicity. In case 3, the photodetector collects signal along $k_x = -\frac{f}{2}$. The total optical field along this direction is: $U_{-\frac{f}{2}} = Ee^{ik_z z}\left[\tilde{t}_0 + \frac{cN_0}{2}e^{-\frac{t}{\tau_r}} + \frac{cN_0}{4}e^{-\frac{t}{\tau_r}}e^{-f^2 Dt}\right]e^{-i\frac{f}{2}x}$. Therefore, the laser intensity in this direction is: $I_{-\frac{f}{2}} = E^2\left[\tilde{t}_0 + \frac{cN_0}{2}e^{-\frac{t}{\tau_r}} + \frac{cN_0}{4}e^{-\frac{t}{\tau_r}}e^{-f^2 Dt}\right]^2$. When there is no pump, the laser intensity for the unexcited sample in the detected direction is: $I_{-\frac{f_0}{2}} = E^2\tilde{t}_0^2$. So the expression of the differential transmission in case 3 is: $\frac{\Delta T}{T_0} = \frac{I_{-\frac{f}{2}} - I_{-\frac{f_0}{2}}}{I_{-\frac{f_0}{2}}} = \left[1 + e^{-\frac{t}{\tau_r}}\left(\frac{a}{2} + \frac{a}{4}e^{-f^2 Dt}\right)\right]^2 - 1$, where $a\tilde{t}_0 = cN_0 = \Delta\tilde{t}_0$ is the maximum modulation of the amplitude transmission coefficient, with $a$ as the modulation factor. If further considering the carrier cooling effect, the final analytical expression of the signal in case 3 is

$$\frac{\Delta T}{T_0}(t) = \left[1 + \left(A_1 e^{-\frac{t}{\tau_r}} + A_2 e^{-\frac{t}{\tau_{cool}}}\right)\left(\frac{a}{2} + \frac{a}{4}e^{-f^2 Dt}\right)\right]^2 - 1$$

$$\approx a\left(A_1 e^{-\frac{t}{\tau_r}} + A_2 e^{-\frac{t}{\tau_{cool}}}\right)\left(1 + \frac{1}{2}e^{-f^2 Dt}\right) \qquad (3)$$

where $A_1$ and $A_2$ are the amplitudes of carrier recombination and carrier cooling, and the approximation sign holds when $a \ll 1$, which is the case here under the condition of small excitation.

**Case 4: Heterodyne of diffracted detection beam and transmitted reference beam**

As has been discussed before, case 4 and case 3 are symmetric, so the analytical expression of the signal in case 4 should be the same as that of case 3. A derivation similar to case 3 indeed leads to the same result:

$$\frac{\Delta T}{T_0}(t) = \left[1 + \left(A_1 e^{-\frac{t}{\tau_r}} + A_2 e^{-\frac{t}{\tau_{cool}}}\right)\left(\frac{a}{2} + \frac{a}{4}e^{-f^2 Dt}\right)\right]^2 - 1$$

$$\approx a\left(A_1 e^{-\frac{t}{\tau_r}} + A_2 e^{-\frac{t}{\tau_{cool}}}\right)\left(1 + \frac{1}{2}e^{-f^2 Dt}\right) \qquad (4)$$

**Case 5: Diffraction only**

The difference between case 4 and case 5 is that in case 5 there is no reference beam. So only the optical field of the diffracted detection beam is considered, which is $U_{diff} = E e^{ik_z z}\left[\frac{cN_0}{4}e^{-\frac{t}{\tau_r}}e^{-f^2 Dt}\right]e^{i\frac{f}{2}x}$. The diffraction intensity is $I_{diff} = E^2 \left[\frac{cN_0}{4}e^{-\frac{t}{\tau_r}}e^{-f^2 Dt}\right]^2$. When there is no pump laser, the intensity in the diffraction direction is 0. If further considering the carrier cooling effect, the final analytical expression of the diffraction signal in case 5 is

$$I_{diff} = M\left[\left(A_1 e^{-\frac{t}{\tau_r}} + A_2 e^{-\frac{t}{\tau_{cool}}}\right) e^{-f^2 Dt}\right]^2 \tag{5}$$

Equations (1)-(5) are the models for cases 1-5. Parameters $A_1, A_2, \tau_r$, and $\tau_{cool}$ can be obtained from measurements without the amplitude transmission gratings, so the only fitting parameters are the carrier diffusion coefficient $D$ and a scaling factor. Equation (2) and Equations (3) & (4) are exactly the same, which proves that although starting from different perspectives, the grating imaging and the transient grating heterodyne techniques are identical in the final theoretical expression. By fitting the experimental data in Figure 4 with Equations (1)-(5), the carrier diffusion coefficient in each case can be extracted and shown in table 1. All the fitting curves agree with the experimental data very well, suggesting that the derived models well describe each experimental condition. More importantly, the fitted values of diffusion coefficient from all five measurements are consistent, close to an average value of 7.32cm$^2$/s, which falls into the range reported in literature (from several to 50 cm$^2$/s), which depends on different sample structures, growth and doping conditions [2,4,5,18,20]. This fact further validates our derived theoretical models. Our experimental results along with the derived theoretical models demonstrate that the grating imaging and the transient grating heterodyne methods can both be utilized to measure the in-plane carrier diffusion and yield the same results, with signal-to-noise ratio significantly improved compared with pure diffraction.

## V. CONCLUSIONS

In summary, by varying the experimental setup with 5 different geometries, compared are three transient optical grating techniques (transient grating heterodyne, grating imaging and transient grating diffraction) in measuring the carrier diffusion dynamics in a GaAs/AlAs superlattice. Theoretical models are derived for each experimental geometry to extract the in-plane carrier diffusion coefficient. It is demonstrated that, both experimentally and theoretically, the transient grating heterodyne and the grating imaging methods are identical, even though from two different measurment perspectives. The raw signal of the grating imaging (transient grating heterodyne) method is about 315 (157) times larger than that of the transient grating diffraction, manifesting the advantage in SNR of the grating imaging (transient grating heterodyne) technique over the transient grating diffraction. In addition to measure carrier diffusion, the grating imaging and the transient grating heterodyne methods can be utilized to measure in-plane transport of phonon, heat, and electron spin as well.

**Supplemental Information**

Additional supplemental information may be found in the online version of this article at the publisher's website.

**Acknowledgements**

The authors appreciate helpful and stimulating discussions with Dr. Bai Song. The authors would like to acknowledge supports from National Science Foundation (NASCENT, Grant No. EEC-1160494; CAREER, Grant No. CBET-1351881; CBET-1707080); Department of Energy (SBIR/STTR, Grant No. DE-SC0013178); and DOD_ Army (Grant No. W911NF-16-1-0559).

Received:     ((will      be      filled      in      by      the      editorial      staff))
Revised:      ((will      be      filled      in      by      the      editorial      staff))
Published online: ((will be filled in by the editorial staff))

**REFERENCES**

[1] A. Miller, "Transient grating studies of carrier diffusion and mobility in semiconductors," *Nonlinear Optics in Semiconductors II,* vol. 59, pp. 287-312, 1998.

[2] A. Cameron, P. Riblet, and A. Miller, "Spin gratings and the measurement of electron drift mobility in multiple quantum well semiconductors," *Phys. Rev. Lett.,* vol. 76, no. 25, p. 4793, 1996. DOI: 10.1103/PhysRevLett.76.4793

[3] B. A. Ruzicka, L. K. Werake, H. Samassekou, and H. Zhao, "Ambipolar diffusion of photoexcited carriers in bulk GaAs," *Appl. Phys. Lett.,* vol. 97, no. 26, p. 262119, 2010. DOI: 10.1063/1.3533664

[4] H. Zhao, M. Mower, and G. Vignale, "Ambipolar spin diffusion and D'yakonov-Perel'spin relaxation in GaAs quantum wells," *Phys. Rev. B,* vol. 79, no. 11, p. 115321, 2009. DOI: 10.1103/PhysRevB.79.115321

[5] K. Chen *et al.*, "Measurement of ambipolar diffusion coefficient of photoexcited carriers with ultrafast reflective grating-imaging technique," *ACS Photonics,* vol. 4, no. 6, pp. 1440-1446, 2017. DOI: 10.1021/acsphotonics.7b00187

[6] K. Chen *et al.*, "Non-destructive measurement of photoexcited carrier transport in graphene with ultrafast grating imaging technique," *Carbon,* vol. 107, pp. 233-239, 2016. DOI: 10.1016/j.carbon.2016.05.075

[7] P. Voehringer and N. F. Scherer, "Transient grating optical heterodyne detected impulsive stimulated Raman scattering in simple liquids," *J. Phys. Chem.,* vol. 99, no. 9, pp. 2684-2695, 1995. DOI: 10.1021/j100009a027

[8] S. Fujiyoshi, S. Takeuchi, and T. Tahara, "Time-resolved impulsive stimulated Raman scattering from excited-state polyatomic molecules in solution," *J. Phys. Chem. A,* vol. 107, no. 4, pp. 494-500, 2003. DOI: 10.1021/jp0270856

[9] A. Maznev, K. Nelson, and J. Rogers, "Optical heterodyne detection of laser-induced gratings," *Opt. Lett.,* vol. 23, no. 16, pp. 1319-1321, 1998. DOI: 10.1364/OL.23.001319

[10] J. A. Johnson *et al.*, "Phase-controlled, heterodyne laser-induced transient grating measurements of thermal transport properties in opaque material," *J. Appl. Phys.,* vol. 111, no. 2, p. 023503, 2012. DOI: 10.1063/1.3675467

[11] T. Kim, D. Ding, J.-H. Yim, Y.-D. Jho, and A. J. Minnich, "Elastic and thermal properties of free-standing molybdenum disulfide membranes measured using ultrafast transient grating spectroscopy," *APL Materials,* vol. 5, no. 8, p. 086105, 2017. DOI: 10.1063/1.4999225

[12] L. Yang, J. Koralek, J. Orenstein, D. Tibbetts, J. Reno, and M. Lilly, "Measurement of electron-hole friction in an n-doped GaAs/AlGaAs quantum well using optical transient grating spectroscopy," *Phys. Rev. Lett.,* vol. 106, no. 24, p. 247401, 2011. DOI: 10.1103/PhysRevLett.106.247401

[13] K. Chen *et al.*, "Carrier Trapping by Oxygen Impurities in Molybdenum Diselenide," *ACS Appl. Mater. Interfaces,* 2017. DOI: 10.1021/acsami.7b15478

[14] N. Kumar, Q. Cui, F. Ceballos, D. He, Y. Wang, and H. Zhao, "Exciton-exciton annihilation in MoSe$_2$ monolayers," *Phys. Rev. B,* vol. 89, no. 12, p. 125427, 2014. DOI: 10.1103/PhysRevB.89.125427

[15] G. Chen, *Nanoscale energy transport and conversion: a parallel treatment of electrons, molecules, phonons, and photons*. Oxford University Press, 2005.

[16] D. Chemla, D. Miller, P. Smith, A. Gossard, and W. Wiegmann, "Room temperature excitonic nonlinear absorption and refraction in GaAs/AlGaAs multiple quantum well structures," *IEEE J. Quantum Electron.,* vol. 20, no. 3, pp. 265-275, 1984. DOI: 10.1109/JQE.1984.1072393

[17] R. Wang, B. A. Ruzicka, N. Kumar, M. Z. Bellus, H.-Y. Chiu, and H. Zhao, "Ultrafast and spatially resolved studies of charge carriers in atomically thin molybdenum disulfide," *Phys. Rev. B,* vol. 86, no. 4, p. 045406, 2012. DOI: 10.1103/PhysRevB.86.045406

[18] H. Hillmer, A. Forchel, and C. Tu, "Enhancement of electron-hole pair mobilities in thin GaAs/Al$_x$Ga$_{1-x}$As quantum wells," *Phys. Rev. B,* vol. 45, no. 3, p. 1240, 1992. DOI: 10.1103/PhysRevB.45.1240

[19] M. Achermann, B. Nechay, F. Morier-Genoud, A. Schertel, U. Siegner, and U. Keller, "Direct experimental observation of different diffusive transport regimes in semiconductor nanostructures," *Phys. Rev. B,* vol. 60, no. 3, p. 2101, 1999. DOI: 10.1103/PhysRevB.60.2101

[20] H. Akiyama, T. Matsusue, and H. Sakaki, "Carrier scattering and excitonic effects on electron-hole-pair diffusion in nondoped and p-type-modulation-doped GaAs/Al$_x$Ga$_{1-x}$As quantum-well structures," *Phys. Rev. B,* vol. 49, no. 20, p. 14523, 1994. DOI: 10.1103/PhysRevB.49.14523

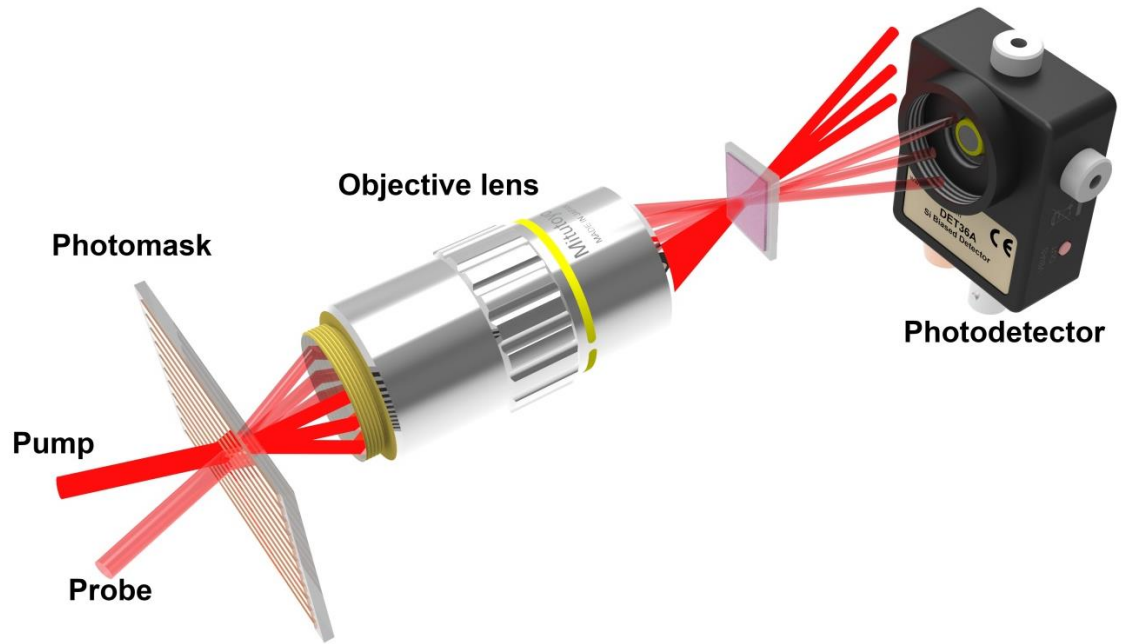

**Figure 1**. Experimental setup of the grating imaging method. Pump and probe pulses are from the same laser source with the same wavelength but separated spatially and delayed temporally.

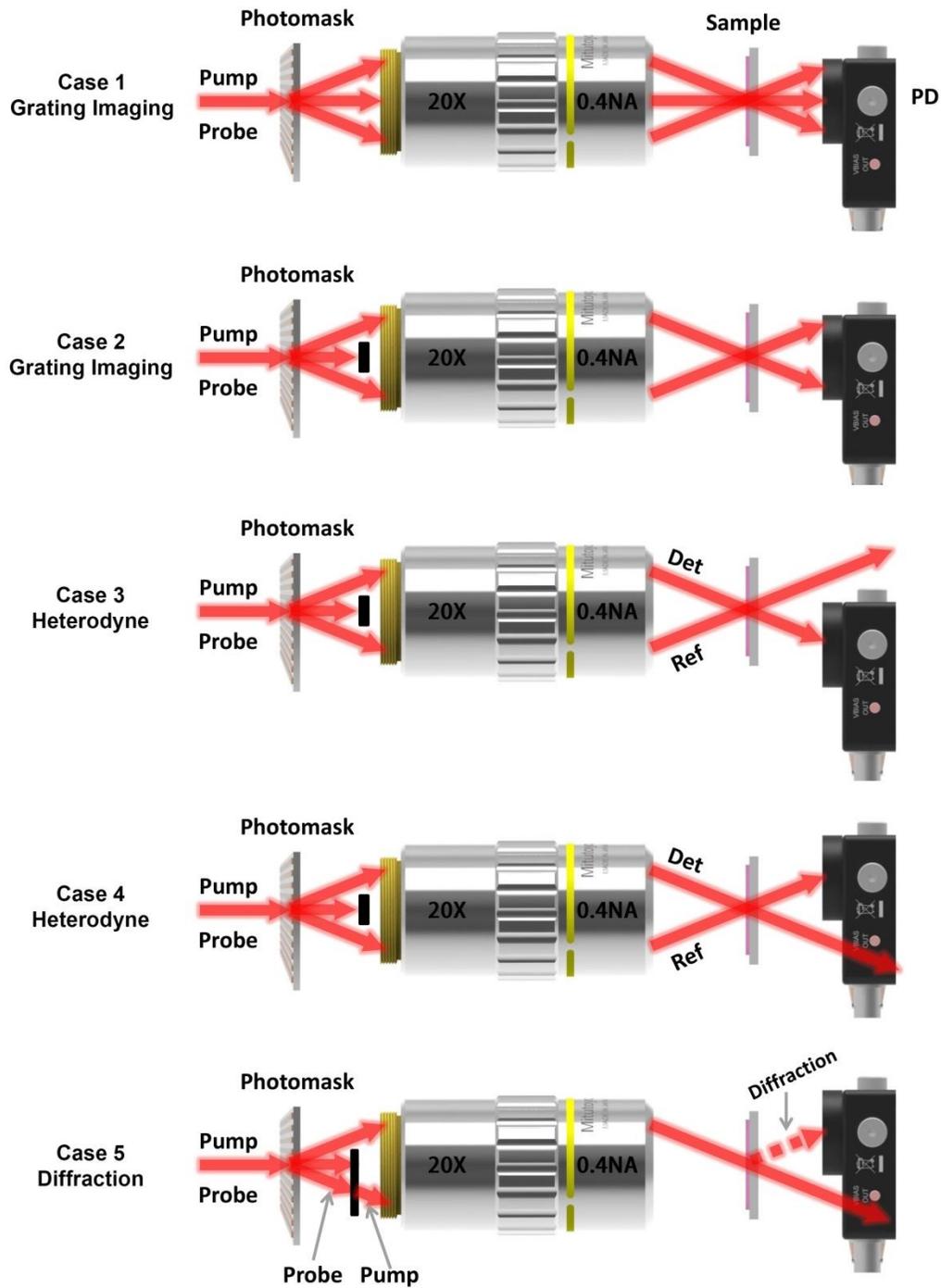

**Figure 2**. Different variations of the grating imaging setup to achieve three different measuring techniques. All the figures are side view. Only the probe related beams are shown at the right side of the objective lens for simplicity. Det and Ref stand for detection and reference beams, respectively.

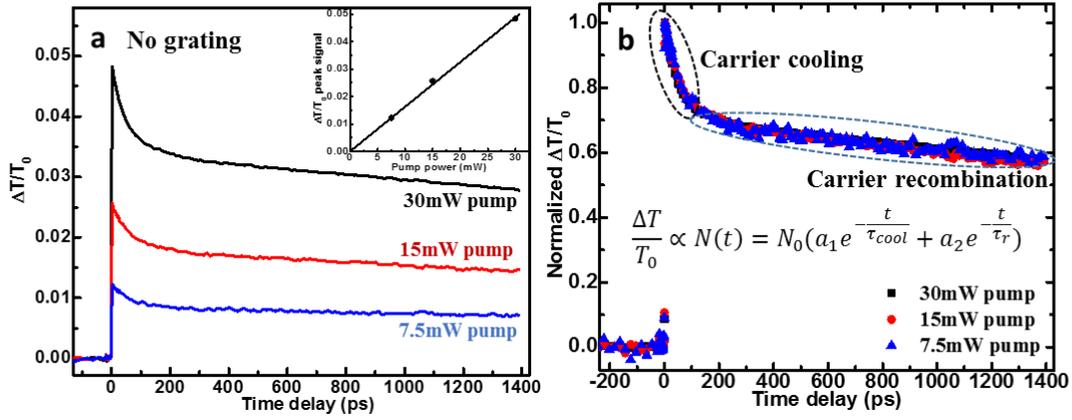

**Figure 3**. (a) Differential transmission signals ΔT/T0 of the GaAs (6nm)/AlAs (6nm) superlattice measured at different pump powers. Inset: Peak ΔT/T$_0$ signals vs pump power. (b) Normalized ΔT/T$_0$ signals at different pump powers.

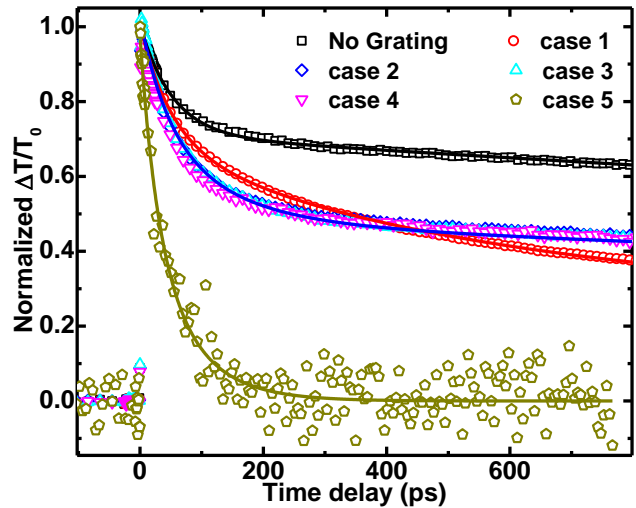

**Figure 4**. Normalized ΔT/T$_0$ signals of different cases indicated in Figure 2. The solid lines are the fitting curves using the theoretical models derived in section IV.

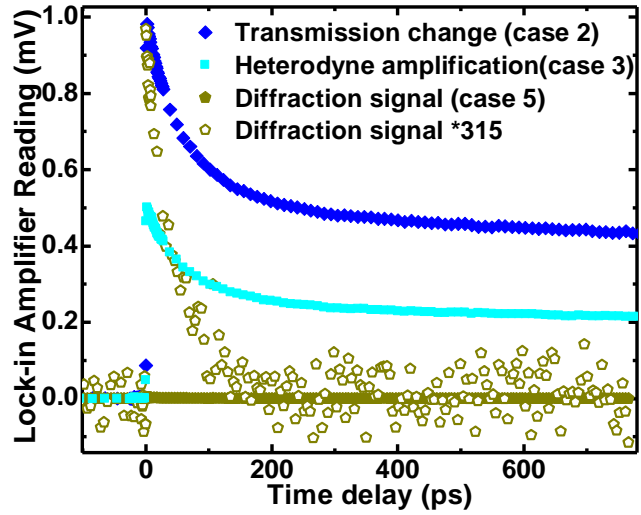

**Figure 5**. Comparison of the raw signals (LIA reading) among the transmission change (case 2), the heterodyned diffraction (case 3), and the pure diffraction (case 5) induced by the same carrier density grating.

**Table 1**. The fitted carrier diffusion coefficients for different cases

|  | Case 1 | Case 2 | Case 3 | Case 4 | Case 5 |
|---|---|---|---|---|---|
| $D$ (cm$^2$/s) | 7.35 ±0.05 | 7.45 ±0.4 | 7.42 ±0.5 | 7.42 ±0.5 | 6.95 ±0.4 |

# Supplemental Material

# Comparison between Grating Imaging and Transient Grating Techniques on Measuring Carrier Diffusion in Semiconductor

*Ke Chen[1], Xianghai Meng[1], Feng He[1,2], Yongjian Zhou[1], Jihoon Jeong[1], Nathanial Sheehan[3], Seth R Bank[3], and Yaguo Wang[1,2]**

4. Department of Mechanical Engineering, The University of Texas at Austin, Austin, TX, 78712, USA
5. Texas Materials Institute, The University of Texas at Austin, Austin, TX, 78712, USA
6. Department of Electrical and Computer Engineering, The University of Texas at Austin, Austin, TX, 78758, USA

*Corresponding Author. Email: yaguo.wang@austin.utexas.edu

1. **Photoluminescence of the sample under test: a 30-periods GaAs(6nm)/AlAs(6nm) supperlattice**

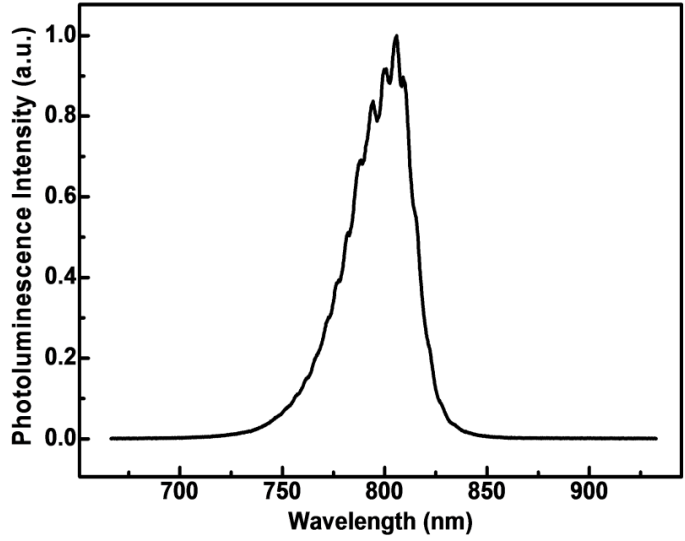

Figure s1. Photoluminescence of the sample under test: 30 periods of GaAs(6nm)/AlAs(6nm) supper lattice

## 2. Analysis on the relative change of both the real and imaginary parts of refractive index induced by carrier excitation

Generally speaking, after a pump pulse generates excited carriers in a semiconductor, both the real part $n$ and the imaginary part $\kappa$ of the refractive index of the material will change, i.e. $n=n_0+\Delta n$, $\kappa=\kappa_0+\Delta\kappa$, which will result in a change in both reflectivity and transmittance, $R=R_0+\Delta R$, and $T=T_0+\Delta T$. However, the excited-carrier-induced changes in $n$ and $k$ and their importance will be different in different detection regions. In the following analysis, we will show that in resonant detection region where the probe photon energy is close to the bandgap of the material, $\Delta\kappa/\kappa_0$ (the absorption change) is typically much larger than $\Delta n/n_0$ (the phase change); but at probe wavelengths far away from resonance, $\Delta\kappa/\kappa_0$ is usually negligible while $\Delta n/n_0$ will dominate.

In a real physical system, the $\Delta n/n_0$ and $\Delta\kappa/\kappa_0$ are not independent parameters, but correlated to each other through the Kramers-Kronig relation, which is expressed as:

$$n(\omega) = 1 + \frac{2}{\pi} P \int_0^\infty \frac{\omega' \kappa(\omega')}{\omega'^2 - \omega^2} d\omega', \quad (s1)$$

$$\kappa(\omega) = \frac{2\omega}{\pi} P \int_0^\infty \frac{n(\omega')-1}{\omega'^2 - \omega^2} d\omega', \quad (s2)$$

Hence, the change of the real and imaginary parts of refractive index are also correlated:

$$\Delta n(\omega) = \frac{2}{\pi} P \int_0^\infty \frac{\omega' \Delta\kappa(\omega')}{\omega'^2 - \omega^2} d\omega', \quad (s3)$$

$$\Delta\kappa(\omega) = \frac{2\omega}{\pi} P \int_0^\infty \frac{\Delta n(\omega')}{\omega'^2 - \omega^2} d\omega', \quad (s4)$$

If the magnitude of either $n$ change or $\kappa$ change is known, we can use Equations (s3)~(s4) to estimate the change of the other one. In a semiconductor structure with well-defined band structure, such as GaAs/AlAs supper lattice in our case, the excited carrier induced absorption change ($\Delta\kappa$) is typically due to the phase space filling effect. After carrier thermalization and cooling, the excited carriers will mainly occupy the band edge energy states (or the exciton state for 2D or low temperature cases), giving rise to an absorption change **only non-trivial at around band edge** (or the exciton energy).[1–4] Since photoluminescence (PL) signal can just reflect the distribution of the excited carriers at the band edge, the absorption change -$\Delta\kappa(\omega)$ and the PL signal typically will have the same shape.[1,5] Based on this understanding, we can assume the absorption change with the following expression:

$$\Delta\kappa(\omega) = -A(\omega)\exp\left(-4\ln 2 \frac{(\omega-\omega_g)^2}{\Gamma^2}\right)\kappa_0(\omega), \quad (s5)$$

where $\kappa_0(\omega)$ is the extinction coefficient before excitation, which has been measured and modeled in references,[6,7] $\omega_g$ is the angular frequency corresponding to the band gap of the direct transition, and $\Gamma$ is a line width parameter characterizing the occupied energy range, and $A(\omega)$ is the absorption reduced ratio with a step-function like shape (to eliminate the part with energies lower than the direct band gap where little is contributed to the absorption change). The profile of carrier induced absorption change $\Delta\kappa(\omega)/\kappa_0$ based on equation (s5) is plotted in Figure s2 as the black curve. By substituting Equation (s5) into Equation (s3), we can calculate the correlated change of real part of refractive index, $\Delta n/n_0$, which is also plotted in Figure s2. It can be seen that, in the resonant region (marked by green dashed rectangle), $\Delta\kappa/\kappa_0$ is much larger than $\Delta n/n_0$, showing at the band edge states, the excited carriers have much greater influence on the extinction coefficient (absorption) than on the real refractive index. In the non-resonant regions (marked by blue dashed rectangles), $\Delta\kappa/\kappa_0$ is negligible but $\Delta n/n_0$ is non-trivial, showing the excited carriers mainly cause change in the real part of refractive index in these regions. Physically, the dominant change in $\Delta\kappa/\kappa_0$ at the band edge states comes from the phase-space filling (Pauli blocking effect). Because $\frac{1}{\omega'^2 - \omega^2}$ in equation (s3) has odd symmetry with respect to $\omega$ and vanishes away from $\omega$, and the integrand in equation (s3) at the band edge $\omega$ is a product of the odd symmetric term with a smooth and gradual change term $\omega'\Delta\kappa(\omega')$, hence, in the resonant region, this integrand will have opposite sign with comparable magnitude, which will result in a major cancellation when performing the integration to get the refractive index change $\Delta n(\omega)$. So mathematically it is reasonable to reach a small $\Delta n(\omega)$ around the band edge $\omega$.

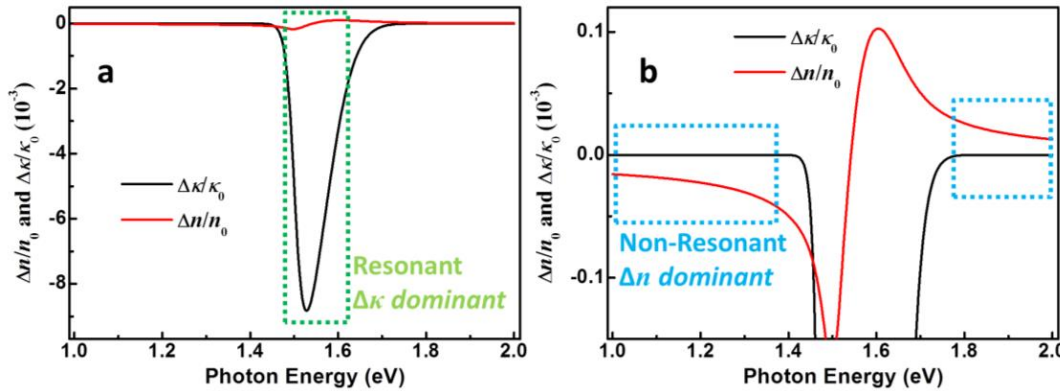

Figure s2. The excited-carrier-induced refractive index change in GaAs/AlAs supperlattice calculated from Kramers-Kronig relation. (a) In the resonant region (circled by green dashed rectangle), $\Delta\kappa/\kappa_0$ is much larger than $\Delta n/n_0$; (b) while in the non-resonant region (circled by blue dashed rectangles), $\Delta\kappa/\kappa_0$ is almost 0 but $\Delta n/n_0$ is non-trivial.

In our experiment, the probe wavelength is 800 nm with the photon energy close to the band gap (1.53 eV, see Figure s1). Therefore, the assumption made in the model derivation in the manuscript that the carrier grating mainly modulates the imaginary part of the refractive index is valid for the probe laser.

References:

(1) Sun, D.; Rao, Y.; Reider, G. A.; Chen, G.; You, Y.; Brézin, L.; Harutyunyan, A. R.; Heinz, T. F. Observation of rapid exciton–exciton annihilation in monolayer molybdenum disulfide. *Nano Lett.* **2014,** *14* (10), 5625-5629.
(2) Bennett, B. R.; Soref, R. A.; Del Alamo, J. A. Carrier-induced change in refractive index of InP, GaAs and InGaAsP. *IEEE J. Quantum Electron.* **1990,** *26* (1), 113-122.
(3) Wang, Y.-T.; Luo, C.-W.; Yabushita, A.; Wu, K.-H.; Kobayashi, T.; Chen, C.-H.; Li, L.-J. Ultrafast multi-level logic gates with spin-valley coupled polarization anisotropy in monolayer $MoS_2$. *Sci. Rep.* **2015,** *5*.
(4) Steinhoff, A.; Rosner, M.; Jahnke, F.; Wehling, T.; Gies, C. Influence of excited carriers on the optical and electronic properties of $MoS_2$. *Nano Lett.* **2014,** *14* (7), 3743-3748.
(5) Kumar, N.; Cui, Q.; Ceballos, F.; He, D.; Wang, Y.; Zhao, H. Exciton-exciton annihilation in $MoSe_2$ monolayers. *Phys. Rev. B* **2014,** *89* (12), 125427.
(6) Aspnes, D. E.; Kelso, S. M.; Logn, R. A.; Bhat, R. Optical properties of $Al_xGa_{1-x}As$. Journal of Applied Physics **1986**, 60, 754
(7) Adachi, S. Optical dispersion relations for GaP, GaAs, GaSb, InP, InAs, InSb, $Al_xGa_{1-x}As$ and $In_{1-x}Ga_xAs_yP_{1-y}$. Journal of Applied Physics **1989**, 66, 6030